\documentclass[twocolumn,showpacs,preprintnumbers,aps,superscriptaddress]{revtex4}

\usepackage{graphicx}
\usepackage{amsmath}
\usepackage{amssymb}
\usepackage{mathrsfs}
\usepackage[english]{babel}

\usepackage{amsthm}
\newtheorem{proposition}{Proposition}
\newtheorem{lemma}{Lemma}
\newtheorem{theorem}{Theorem}

\begin{document}

\title{Microdynamics and Criticality of Adaptive Regulatory Networks}

\author{Ben D. MacArthur}
\email{ben.macarthur@mssm.edu}
\affiliation{Department of Pharmacology and Systems Therapeutics, Systems Biology Center New York (SBCNY), Mount Sinai School of Medicine, New York, NY, USA.}
\author{Rub\'en J. S\'anchez-Garc\'\i{}a}
\affiliation{Mathematisches Institut, Heinrich-Heine Universit\"at D\"usseldorf, Universit\"atsstr 1, 40225, D\"usseldorf, Germany.}
\author{Avi Ma'ayan}
\affiliation{Department of Pharmacology and Systems Therapeutics, Systems Biology Center New York (SBCNY), Mount Sinai School of Medicine, New York, NY, USA.}

\date{\today}

\begin{abstract}
We present a model of adaptive regulatory networks consisting of a simple biologically-motivated rewiring procedure coupled to an elementary stability criterion. The resulting networks exhibit a characteristic stationary heavy-tailed degree distribution, show complex structural microdynamics and self-organize to a dynamically critical state. We show analytically that the observed criticality results from the formation and breaking of transient feedback loops during the adaptive process.
\end{abstract}


\maketitle

\emph{Introduction}.-- Much recent research attention has focused on understanding the structure of naturally occurring empirical networks and associated random graph models \cite{albert, newman}. An overarching aim of many of these studies is to determine the relationships between network structure and dynamics. For instance, the presence of modularity and sparsity have long been known to contribute to global stability, while the presence of feedback is a well-studied prerequisite for the support of complex dynamics such as oscillations, multistability and chaos \cite{may, thomas}. However, although much work has, so far, focused on networks which are static in their topology, many real-world complex systems evolve both structurally \emph{and} dynamically over time \cite{batty, luscombe, gautreau}. For instance, neural networks change in structure depending on synaptic activity while genetic regulatory networks change structurally on the evolutionary time scale in a fitness-dependent manner. Consequently, \emph{adaptive} networks -- in which changes in network topology and dynamics continually feedback on each other -- are now attracting increasing research interest \cite{gross}. Since many biological regulatory systems, such as neural and genetic regulatory networks, are also thought to optimally balance stability and adaptability by operating at, or near to, criticality \cite{beggs,shmulevich,nykter,balleza}, a number of prior studies have sought to elucidate mechanisms of self-organized criticality (SOC) in adaptive networks \cite{bornholdta, luque, bornholdt2, liu, garlaschelli, rohlf, magnasco, meisel}. For example, important early results were obtained by Christensen \emph{et al.} and Bornholdt and Rohlf, who showed that adaptive networks may self-organize to a critical state by a simple mechanism in which `\emph{quiet nodes grow links} [\emph{and}] \emph{active nodes lose links}' \cite{christensen, bornholdt}. However, despite the apparent ubiquity of critical adaptive networks in nature, the mechanisms of adaptive SOC remain to be fully determined. 

In this article we outline a simple new adaptive network model which reproduces characteristic features of biological systems, including a heavy-tailed degree distribution and self-organization to a dynamically critical state. To fix ideas our model may be thought of as describing adaptive changes in a genetic regulatory network, although the model may also be applied more generally to other systems which undergo adaptive rewiring. In genetic regulatory networks, genetic mutations cause changes in protein structure which, in turn, not only alter \emph{local} network connectivity but also \emph{global} system stability. Consequently, our model is a simple scheme intended to describe, albeit in a highly idealized way, mutation-driven local rewiring in the face of a global stability (fitness) constraint: mutations are allowed to accumulate during times of stability, but harmful mutations are suppressed during times of instability. 

\emph{Preliminaries}.-- Mathematically a network is a graph consisting of a set of vertices (or nodes) $V$ of size $n$ and a set of edges (or links) $E$. A directed graph (digraph) is a graph in which each edge $v_i \sim v_j \in E$ has a unique orientation ($v_i \to v_j$). Although digraphs describe well structural relationships in complex systems, in many cases relationships also have an intrinsic sign -- friendship and enmity in social networks or activation and inhibition in biochemical regulatory networks, for instance. To cope with such systems, a natural framework is that of signed digraphs. A signed digraph $\vec{S}$ is a digraph in which each edge $v_i \sim v_j \in E$ additionally has a unique sign $\sigma_{ij} \in [-1,+1]$ depending on whether it is `activating' ($\sigma_{ij} = +1$) or `inhibiting' ($\sigma_{ij} = -1$). The adjacency matrix $\mathbf{A} = a_{ij}$ of a signed digraph has the form $a_{ij} = \sigma_{ij}$ if $v_i \sim v_j \in E$ and $a_{ij} = 0$ otherwise. When considering structural features of $\vec{S}$ without regard for signs we shall also make use of the absolute adjacency matrix $\tilde{\mathbf{A}} = \vert a_{ij} \vert$. The in-degree (out-degree) of a vertex is the number of in-coming (out-going) edges it has, without regard for sign. The net-degree $d_{\textrm{net}}(v_i) = \vert d_{\textrm{in}}(v_i) - d_{\textrm{out}}(v_i) \vert$ of a vertex $v_i$ as the absolute difference of its in-coming and out-going degree. Intuitively, net-degree measures how `source-' or `sink'-like a vertex is. By extension, we define the \emph{imbalance} of a vertex-pair as the absolute difference of their net-degrees, $\textrm{I}(v_i, v_v) = \vert d_{\textrm{net}}(v_i) - d_{\textrm{net}}(v_j) \vert$. It has recently been observed that many empirical networks contain significantly more source- and sink-vertices than expected by chance, and that this degree imbalance naturally leads to depletion of feedback loops (cycles) which, in turn, confers enhanced stability properties \cite{maayan}. Thus, degree imbalance and dynamic stability are intrinsically related, a fact that our model exploits. 

\emph{Model}.-- We begin at $t = 0$ with a random signed digraph $\vec{S}(t=0)$ of size $n$ with Erd\H{o}s-R\'{e}nyi connectivity, in which edge orientations and signs have been assigned independently in an equiprobable random manner \footnote{For efficiency in the simulations shown we set the edge-inclusion probability $\approx \ln (n) /n$ and consider a maximally sparse connected random graph. Qualitatively the same results may be achieved for more dense graphs.}. We then rewire $\vec{S}(t)$ at successive time-steps according to the following rules: (1) randomly and uniformly chose an edge $e_{\textrm{old}} = v_a \sim v_b$ connecting two vertices in $\vec{S}(t)$ such that $\vec{S}(t) - e_{\textrm{old}}$ is not disconnected and an ordered pair of non-adjacent vertices $v_c,v_d \neq v_c$. (2) Calculate the pair-wise imbalances $\textrm{I}(v_a,v_b)$ and $\textrm{I}(v_c,v_d)$. (3) Delete $e_{\textrm{old}}$ and create a new edge $e_{\textrm{new}} = v_c \to v_d$, choosing its sign randomly and uniformly, and recalculate the imbalances. (4) If the sum of the two imbalances after the switch is greater than that before then accept the switch unconditionally, otherwise accept with probability $\rho(t)$. 

In order to couple structural rearrangement to dynamics we allow $\rho(t)$ to vary in a manner which takes into account the changing stability of the system. To do so we assume that, in addition to regulatory links defined by $\vec{S}(t)$, each species (vertex $v_i$) also decays at a constant characteristic rate $\epsilon_i$ which we fix at $t = 0$ independently, randomly and uniformly on the unit interval. Thus, at each evolutionary time-point we obtain a modified adjacency matrix $\mathbf{B}(t) = \mathbf{A}(t) - \textrm{diag}(\epsilon_i)$. Global stability is then given by the magnitude of $\mu_{\textrm{max}}(t) = \textrm{max } \textrm{Re } \mu_i(t) $, where $\mu_i(t)$ for $i = 1 \ldots n$ are the eigenvalues of $\mathbf{B}(t)$. In particular, the system is stable when $\mu_{\textrm{max}}(t) < 0$ and unstable when $\mu_{\textrm{max}}(t) > 0$ \footnote{This stability criterion assumes that $\mathbf{B}(t)$ is the Jacobian matrix of a dynamical system evaluated at a fixed-point}. Therefore we set $\rho(t) = 1 - h [\mu_{\textrm{max}}(t) ]$, where $h[x]$ is the Heaviside step function, allowing defective switches when the system is stable and suppressing defective switches when the system is unstable. The key property of this coupling is that it makes global information available to the local structural reorganizing process, providing continual feedback between structure and dynamics.    
\begin{figure}[t!] 
\begin{center}
\includegraphics[width=0.49\textwidth]{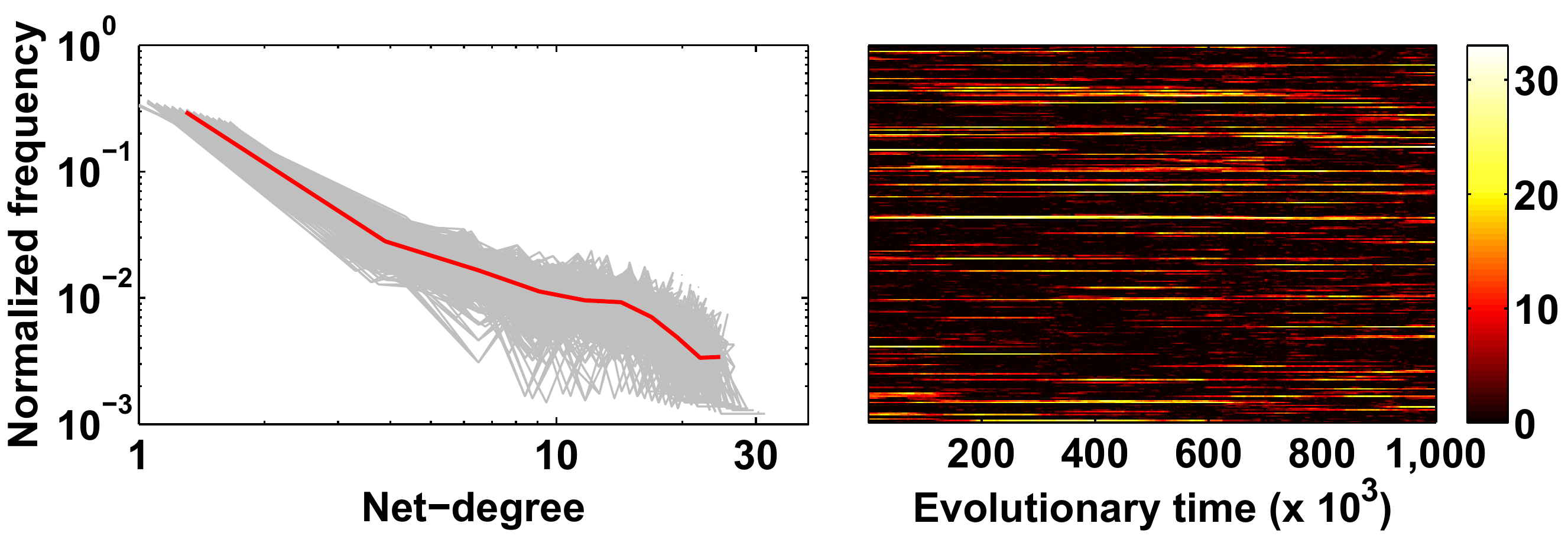}
\caption{\label{heatmaps}\small{Graphs resulting from the evolutionary process exhibit a stationary heavy-tailed degree distribution and complex microdynamics. (\emph{Left}) The net-degree distribution at 1000 time-step intervals for a period of $1 \times 10^6$ time-steps at equilibrium for a typical $250$ vertex network are shown in light gray. In bold (red online) is the mean net-degree distribution over this time period. (\emph{Right}) Each row shows the changing net-degree of one vertex over the same time period.}}
\end{center}
\end{figure}

\emph{Results}.-- The networks produced by this simple model are characterized by a stationary heavy-tailed degree distribution (see Fig.~\ref{heatmaps} left) indicating the presence of hub source- and sink-vertices, a well-known feature of real-world networks \cite{barabasi, maayan}. However, since our model allows for periods of random structural rearrangement, this macroscopic stationarity masks complex structural \emph{microdynamics} in which individual vertices continually accumulate and lose edges and rise and fall in their centrality (see Fig.~\ref{heatmaps} right). This kind of `mixing' microdynamics is not produced by classical rich-get-richer models of hub formation \cite{barabasi}, but has recently been highlighted as an important characteristic of real-world evolving (macroscopically stationary) complex networks \cite{batty, luscombe, gautreau}. 

\begin{figure}[t!] 
\begin{center}
\includegraphics[width=0.45\textwidth]{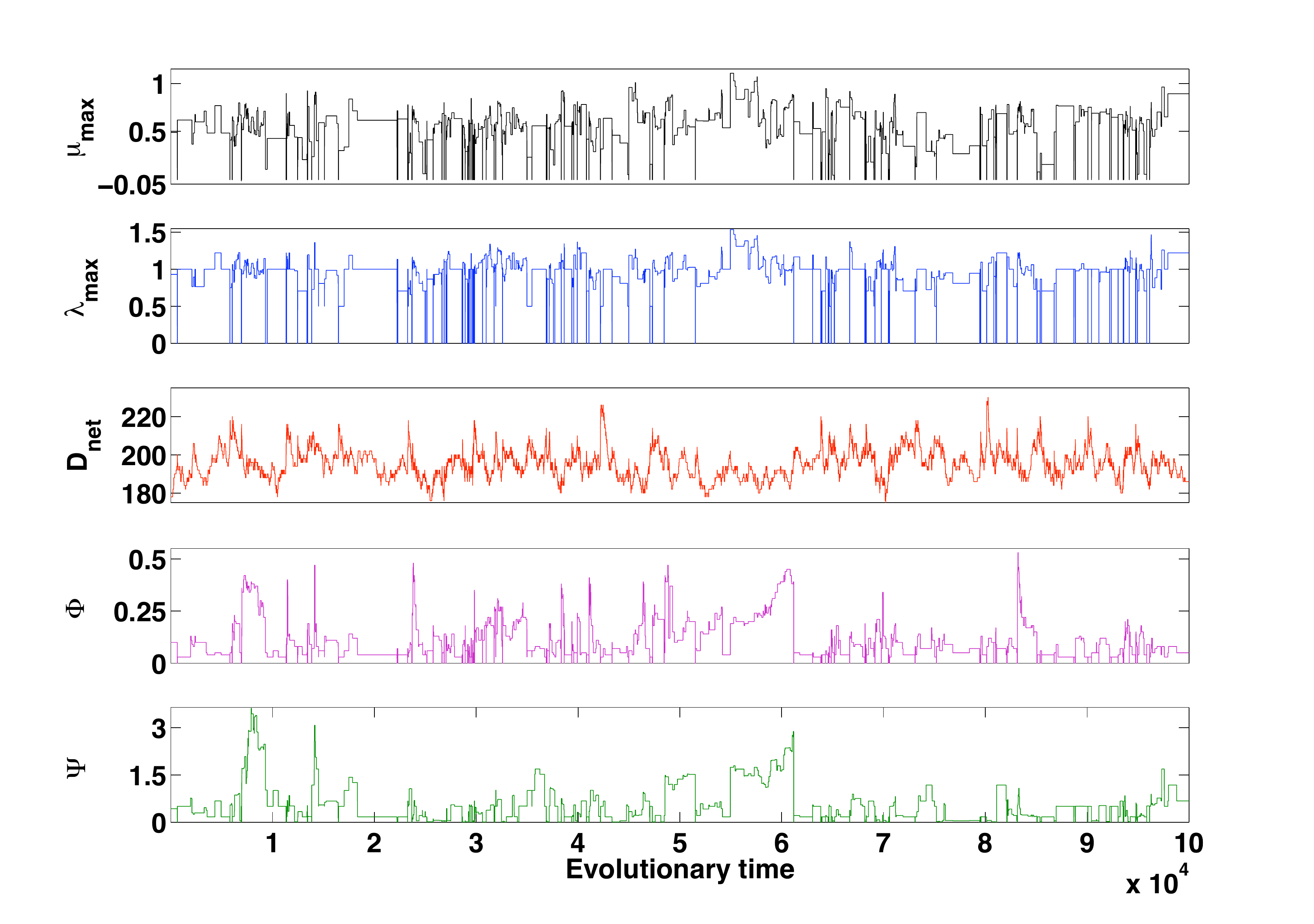}
\caption{\label{threetimeseries}\small{Evolutionary dynamics are characterized by punctuated equilibrium. Dynamics of a representative $100$ vertex network are given. (\emph{Top}) shows the time-series for $\mu_{\textrm{max}}(t)$. Note that $\mu_{\textrm{max}}(t) = - \textrm{min} \, \epsilon_i < 0$ when $\vec{S}(t)$ is acyclic (in this case $\textrm{min} \, \epsilon_i = 6 \times 10^{-3}$); (\emph{Second}) the time-series for $\lambda_{\textrm{max}}(t)$, note the prevalence of $0$ and $+1$ (and less obviously to the eye, but still present, $\cos(\pi/l)$ for some $l \in \mathbb{Z}^+$) in this series as predicted analytically; (\emph{Third}) the total net-degree $\textrm{D}_{\textrm{net}}(t) = \sum_i d_{\textrm{net}}[v_i(t)]$; (\emph{Fourth}) the cyclic index $\Phi(t)$; (\emph{Bottom}) the cyclic index $\Psi(t)$.}}
\end{center}
\end{figure}

Fig.~\ref{threetimeseries} gives a plot of $\mu_{\textrm{max}}(t)$ at equilibrium \footnote{That is, after an initial transient `settling-down' period ($4 \times 10^6$ time-steps prior to data shown).} for a representative system showing that dynamics on the evolutionary time-scale are characterized by periods of stability ($\mu_{\textrm{max}}(t) < 0$) and instability ($\mu_{\textrm{max}}(t) > 0$) punctuating back-and-forth. To help interpret these dynamics, also shown is $\lambda_{\textrm{max}}(t) = \textrm{max } \textrm{Re } \lambda_i(t)$, where $\lambda_i(t)$ are the eigenvalues of the graph adjacency matrix $\mathbf{A}(t)$ and three measures of network structure. The first structural measure shown is total net-degree $\textrm{D}_{\textrm{net}}(t) = \sum_i d_{\textrm{net}} [ v_i(t) ]$, a measure of overall degree imbalance in $\vec{S}(t)$. It is apparent that changes in total net-degree correlate poorly with changes in stability, suggesting that although fluctuations in net-degree are observed during the evolutionary process, it is not degree-imbalance \emph{per se} that drives the characteristic dynamics of $\mu_{\textrm{max}}(t)$. In order to identify more precisely the structural origin of the observed bursting dynamics, and based upon the observation that degree imbalance naturally leads to feedback loop depletion \cite{maayan}, also shown are two measures of network cyclic structure \footnote{A cycle of length $k$ is a non-intersecting path of length $k$ from a vertex back to itself respecting edge directions.}. The first, $\Phi(t)  = n_{\textrm{cyc}}(t)/n$ where $n_{\textrm{cyc}}(t)$ is the number of vertices which participate in a cycle in $\vec{S}(t)$, measures overall cyclic structure without regard for details such as cycle numbers or distribution of cycle lengths. The second,
\begin{equation} \label{eind}
\Psi(t)  = \textrm{Trace }  \textrm{e}^{\tilde{\mathbf{A}}(t)}   - n = \sum_{i=1}^n \textrm{e}^{\tilde{\lambda}_i(t)} - n,
\end{equation}
where $\tilde{\lambda_i}$ are the eigenvalues of the absolute adjacency matrix $\tilde{\mathbf{A}}(t)$, is an indirect measure of `returnability' which takes into account details of closed walks in $\vec{S}(t)$. In particular, $\Psi(t)$ is a sum of all closed walks in $\vec{S}(t)$ weighted in decreasing order by length (note that $\Psi(t) + n$ may be thought of as the partition function of $\vec{S}(t)$) \cite{estrada3, estrada2}.

\begin{figure}[t!] 
\begin{center}
\includegraphics[width=0.45\textwidth]{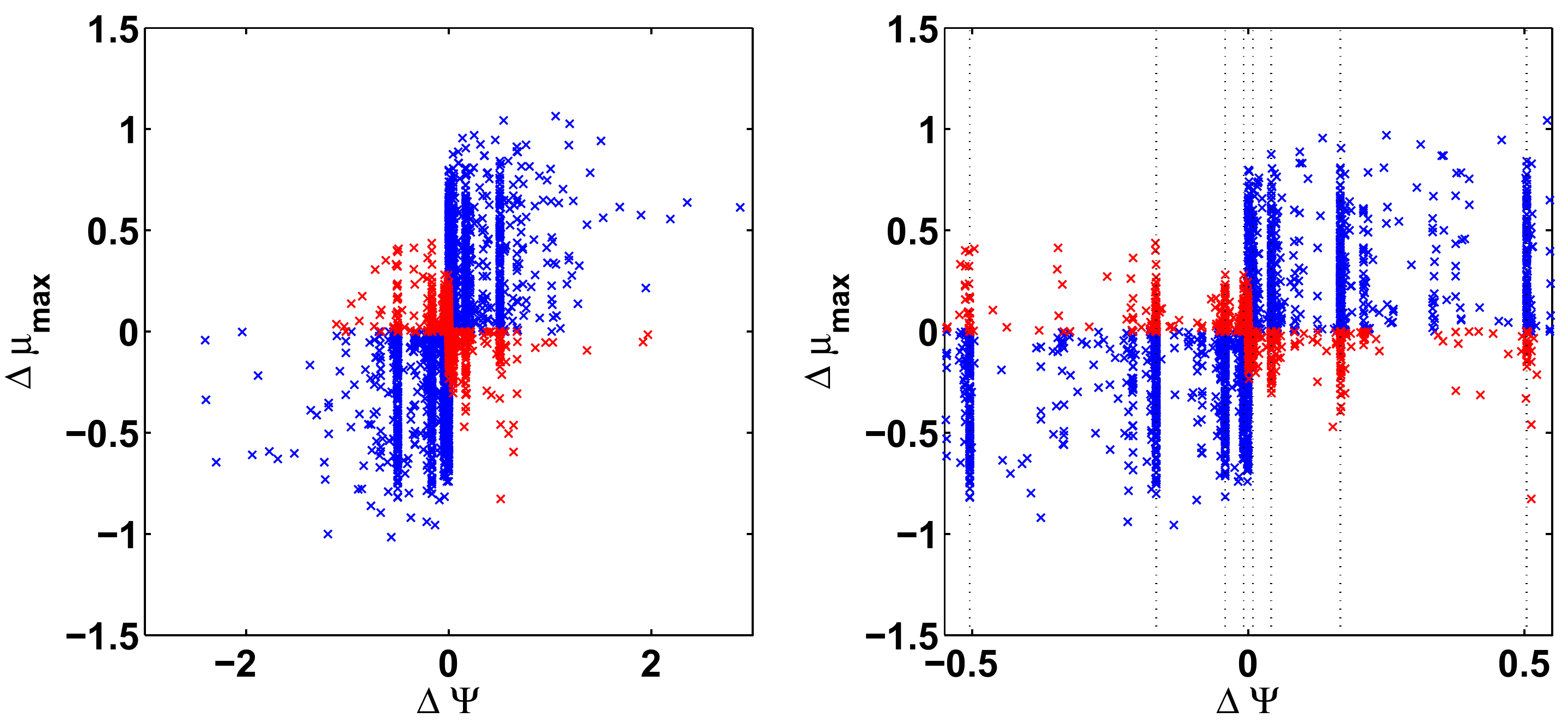}
\caption{\label{scatter1}\small{Transient cycles trigger bursts of instability. (\emph{Left}) A plot of $\Delta \Psi(t) = \Psi (t) - \Psi(t-1)$ against $\Delta \mu_{\textrm{max}}(t) = \mu_{\textrm{max}}(t) - \mu_{\textrm{max}} (t -1)$ using the same data as Fig.~\ref{threetimeseries}. For clarity, the few changes in stability which do not occur concordantly with changes in cyclic structure are shown in light gray (red online). (\emph{Right}) A close-up of the left panel. The striations arise since it is common for isolated cycles to be created or broken during the evolutionary process, triggering changes in stability. The dotted vertical lines are at $\pm \Psi$ calculated analytically using Eq.~\ref{eqnforpsi} with $k = 3 \ldots 6$ and $c_k = 1$.}}
\end{center}
\end{figure}

Examining the time-series of $\Psi(t)$ and $\Phi(t)$ it is apparent that, unlike total net-degree, both $\Psi(t)$ and $\Phi(t)$ exhibit similar bursting behavior to that of $\mu_{\textrm{max}}(t)$. In particular, periods of stability ($\mu_{\textrm{max}}(t) < 0$) generally correspond to periods when both $\Psi(t) = 0$ and $\Phi(t) = 0$ (in Fig.~\ref{threetimeseries} this occurs $>90\%$ of the time). Since $\Psi(t) = 0$ and $\Phi(t) = 0$ if and only if $\vec{S}(t)$ is acyclic this indicates that periods of stability occur primarily when $\vec{S}(t)$ is acyclic. Furthermore, changes in stability predominantly occur concordantly with changes in $\Psi(t)$ and $\Phi(t)$ (for instance, in Fig.~\ref{threetimeseries} this occurs $>99\%$ of the time). Considering the time-series as binary variables (`stable or unstable' and `cyclic or acyclic') and calculating entropies gives $H[\Phi(t)] = H[\Psi(t)] = 0.093$ and $H[\mu_{\textrm{max}}(t)] = 0.100$. The mutual information between these series is $M[\mu_{\textrm{max}}(t),\Phi(t)] = M[\mu_{\textrm{max}}(t),\Psi(t)] = 0.088$, indicating that changes in stability are strongly, although not exclusively, related to changes in cyclic structure (see also Fig.~\ref{scatter1}) \footnote{By discretizing the data we are asking: how much does knowing whether the network is cyclic or not tell us about whether the system is stable or not?}.  

In order to better understand this relationship, we now derive some analytical results relating cycles and spectra of signed digraphs which will help interpret these numerics. To obtain exact results we shall focus on deriving analytic formulae for $\Psi$ and $\lambda_{\textrm{max}}$ in the particular case that all cycles in $\vec{S}$ are disjoint (that is, each vertex $v \in V$ belongs to at most one cycle). Although this is a strong condition to impose, and most real-world networks are not expected to be cycle-disjoint, this case is analytically tractable and, since our evolutionary scheme favors the minimization of cycles, yields results which shed light on the observed dynamics. Full proofs of all analytic results are provided in Appendix \ref{appendixA}.

Firstly we observe that if a signed digraph $\vec{S}$ is cycle-disjoint, then its spectrum has a particularly simple form. Specifically, if $\vec{S}$ contains $c_k^+$ positive cycles and $c_k^-$ negative cycles of length $k$ (for $k = 3 \ldots n$) \footnote{A cycle $c$ is positive (negative) if the product of the edge signs in $c$ equals $+1$ ($-1$).} and all cycles are disjoint, then its spectrum is the zero eigenvalue with multiplicity $(n - n_{\textrm{cyc}})$, along with the eigenvalues of each of the cycles considered separately as induced subgraphs (that is, the union of $c_k^+$ copies of the $k$-th roots of $+1$, and $c_k^-$ copies of the $k$-th roots of $-1$, for $k = 3 \ldots n$). 

An immediate consequence of this result is that if $\vec{S}$ is cycle-disjoint and possesses at least one positive cycle then $\lambda_{\textrm{max}} = 1$, while if all cycles are negative then $\lambda_{\textrm{max}} = \textrm{Re } \textrm{e}^{\pi i/l}  = \cos ( \pi / l)$, where $l$ is the length of the longest cycle in $\vec{S}$. In this sense positive cycles are uniformly destabilizing, while the destabilizing effect of negative cycles increases with length. Examination of the time-series data shows that $\lambda_{\textrm{max}}(t) = 0,1$ and $\cos ( \pi / l)$ for some $ 3 \leq l \leq n \in \mathbb{Z}^+$, do indeed occur commonly during evolution (for instance, in Fig.~\ref{threetimeseries} this occurs $\approx 53 \%$ of the time), indicating the continual formation and breaking of isolated cycles by the evolutionary scheme.

This result is also useful since it allows us to calculate $\Psi$ analytically in the case that $\vec{S}$ is cycle-disjoint. In particular, if $\vec{S}$ contains $c_k$ $(= c_k^+ + c_k^-)$ disjoint cycles of length $k$ for $k = 3 \ldots n$ then, using Eq.~\ref{eind}, 
\begin{equation} \label{eqnforpsi}
\Psi = \sum_{k \, = 3}^{n} c_k \, H_{k,0}(1) - n_{\textrm{cyc}},
\end{equation}
where $H_{k,0}(z)$ is the generalized hyperbolic function of order $k$ and kind $0$ \cite{ungar}. Fig.~\ref{scatter1} shows that values of $\Psi(t)$ calculated using Eq.~\ref{eqnforpsi} often occur during evolution, again indicating that isolated cycles are continually formed and broken by the evolutionary scheme.

\begin{figure}[t!] 
\begin{center}
\includegraphics[width=0.45\textwidth]{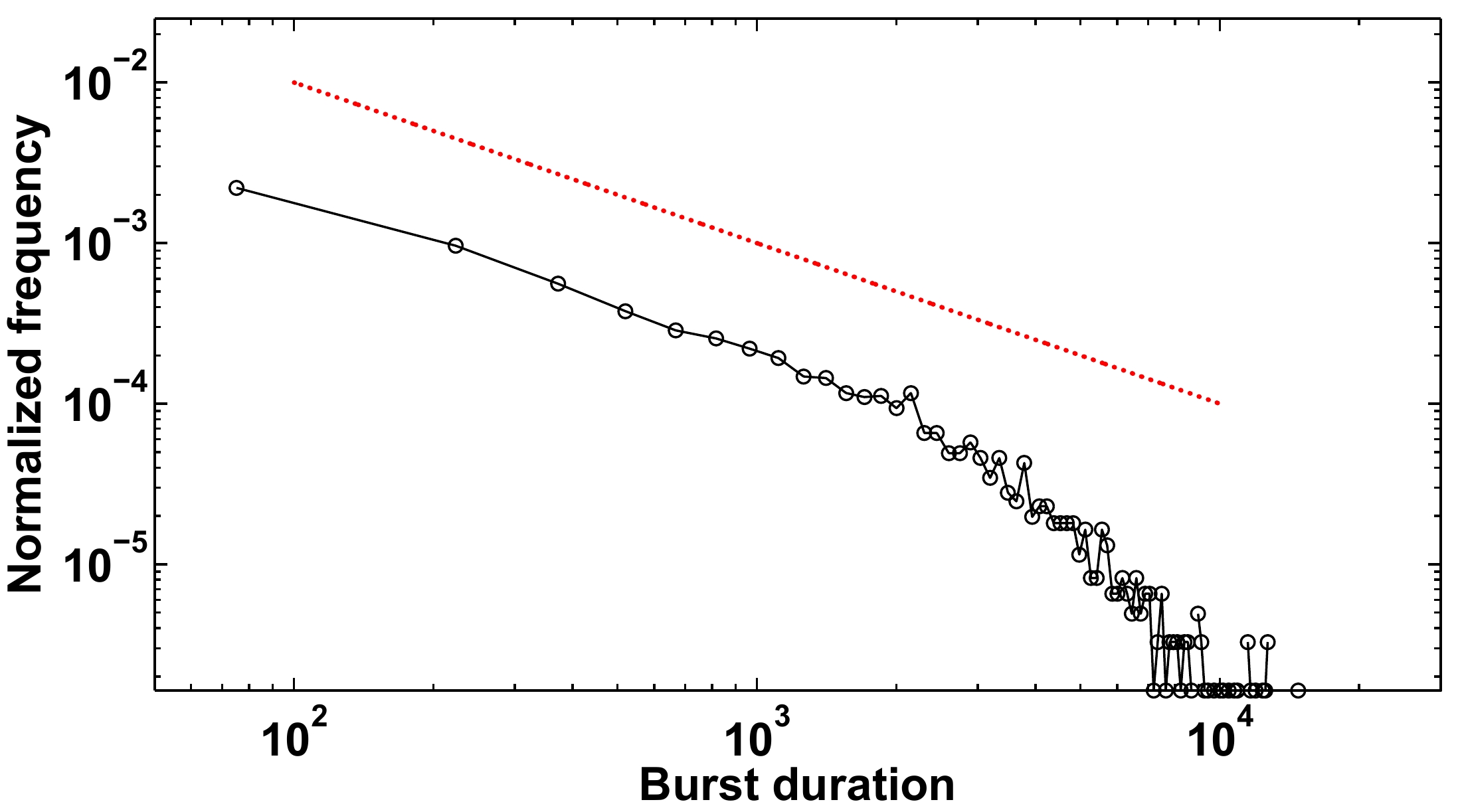}
\caption{\label{powerlaw}\small{Bursts of instability have heavy-tailed statistics. The distribution of burst durations is shown over an interval of $4 \times 10^6$ evolutionary time-steps for a representative system with $100$ vertices at equilibrium. A power-law with exponent $1$ is also shown for reference.}}
\end{center}
\end{figure}

These analytical results may be used to interpret numerics by making use of two further results which relate $\lambda_{\textrm{max}}$ to $\mu_{\textrm{max}}$ in the cycle-disjoint case. Firstly, note that in the special case that $\vec{S}(t)$ is acyclic then $\lambda_{\textrm{max}}(t) = 0$ and $\mu_{\textrm{max}}(t) = - \textrm{min} \, \epsilon_i < 0$, and the system is stable. Secondly, if $\vec{S}(t)$ is cycle-disjoint then $\mu_{\textrm{max}}(t) < \lambda_{\textrm{max}}(t)$ and this bound is tight ($\mu_{\textrm{max}}(t) \to \lambda_{\textrm{max}}(t)$ as $\epsilon_i \to 0$ for all $i$). Consequently, if vertex decay rates are all small then the completion of a single cycle in an otherwise acyclic network is sufficient to trigger a burst of instability, as seen in Fig.~\ref{threetimeseries}. When this occurs the evolutionary process responds by suppressing any further defective switches and rearranging local network structure to remove the cause of the instability. Typically, this is quickly achieved and the burst of instability is relatively short. However, occasionally cycles may accumulate more rapidly than they are removed, giving rise to extended bursts of instability and heavy-tailed statistics characteristic of a critical state (see Fig.~\ref{powerlaw}).

For completeness it should be noted that if $\vec{S}(t)$ is not cycle-disjoint then the relationship between cycles and stability can be complex: it is not necessarily true that $\mu_{\textrm{max}}(t) < \lambda_{\textrm{max}}(t)$ and, in rare cases, changes in cyclic structure and stability may occur discordantly. Further details of when this occurs and a simple worked-example are included in Appendix \ref{appendixB}. 

\emph{Conclusions}.-- Many biological regulatory systems are thought to balance stability and adaptability by self-organizing to a dynamically critical state \cite{beggs,shmulevich,nykter,balleza}. In this article we have presented a simple adaptive network model which reproduces characteristic features of biological systems, including a heavy-tailed connectivity distribution, microdynamics and robust self-organization to criticality. Previous models have shown that adaptive networks may self-organize to a critical state due to rewiring based upon local activity \cite{bornholdta, luque, bornholdt2, liu, garlaschelli, rohlf, magnasco, meisel, christensen, bornholdt}. Here, the mechanism of self-organization is somewhat different and relies on the fact that feedback and stability are generally inversely related: by employing a flexible rewiring scheme which allows feedback loops to be formed during periods of stability and eliminated during periods of instability, criticality naturally arises in our model. It seems plausible that these (and other, as yet unknown) adaptive processes may be responsible for the criticality observed in nature.

\appendix
\section{Proofs}\label{appendixA}
Formal statements and proofs of results stated without proof in the main text are provided here. 

Let $\vec{S} = \vec{S}(V,E)$ be a directed signed graph with vertex-set $V$ of size $n$, edge-set $E$ of size $m$, no dual-edges (if $v_iv_j \in E$ then $v_jv_i \notin E$ for all $i,j = 1 \ldots n$) and no self-loops ($v_iv_i \notin E$ for all $i = 1 \ldots n$). We say that  $\vec{S}$ is \emph{cycle-disjoint} if all cycles in $\vec{S}$ are pair-wise disjoint, that is, if every vertex participates in at most one cycle. Let $c_k^+$ (respectively $c_k^-$) be the number of positive (respectively negative) cycles of length $k$ in $\vec{S}$ for $k = 3 \ldots n$. Thus, the number of vertices which participate in a cycle in $\vec{S}$ is $n_{\textrm{cyc}} = \sum_k k(c_k^+ + c_k^-)$.

\begin{proposition}\label{prop:spectrum}
The eigenvalue spectrum of a cycle-disjoint directed signed graph $\vec{S}$ 
consists of the zero eigenvalue with multiplicity $(n - n_{\textrm{cyc}})$ along with the union of $c_k^+$ copies of the $k$-th roots of unity and $c_k^-$ copies the $k$-th roots of $-1$ for $k = 3 \ldots n$.
\end{proposition}

\begin{proof}
The proof makes use of Sachs' (coefficients) theorem (Theorem 1.32, p32 in \cite{cvetkovic}) which, for completeness, we state here in its general form. 

\begin{theorem}[Sachs]
Let $\vec{W}$ be a weighted digraph with characteristic polynomial $z ^n + a_1 z^{n-1} + \ldots + a_{n-1} z + a_n$ then 
\begin{equation}\label{Sach}
a_i = \sum_{L \in \mathcal{L}_i} (-1)^{p(L)}W(L),
\end{equation}
where $\mathcal{L}_i$ is the set of directed linear subgraphs of $\vec{W}$ on $i$ vertices, $p(L)$ is the number of disjoint components in a given linear subgraph $L$ and $W(L)$ is the product of edge-weights over all edges in $L$. 
\end{theorem}

We now begin our proof of Proposition \ref{prop:spectrum}. Let $\mathbf{A}$ be the adjacency matrix of a cycle-disjoint signed digraph $\vec{S}$. The eigenvalues of $\vec{S}$ are the solutions to the characteristic polynomial of $\mathbf{A}$
\begin{equation} \label{aeval}
z^n + a_1 z ^{n-1} + \ldots + a_{n-1} z + a_n. 
\end{equation}
The largest linear subgraph $L_\textrm{max}$ in $\vec{S}$ consists of the disjoint union of all the cycles in $\vec{S}$ and so has size $n_{\textrm{cyc}} = \sum_k k(c_k^+ + c_k^-)$. Therefore, by Sachs' theorem $a_i = 0$ for $i > n_{\textrm{cyc}}$ and the eigenvalues of $\vec{S}$ are solutions to  
\begin{eqnarray}
z ^n + a_1 z ^{n-1} + \ldots + a_{n_{\textrm{cyc}}} z ^ {n - n_{\textrm{cyc}}} &=& 0 \\ 
z ^{n-n_{\textrm{cyc}}} ( z ^{n_{\textrm{cyc}}} + a_1 z ^{n_{\textrm{cyc}} - 1} + \ldots + a_{n_{\textrm{cyc}}} ) &=& 0.
\end{eqnarray}
Thus, $\vec{S}$ has zero as an eigenvalue with multiplicity $(n-n_{\textrm{cyc}})$. Now let
\begin{equation} \label{leval}
z^{n_{\textrm{cyc}}} + \tilde{a}_1 z ^{n_{\textrm{cyc}} - 1} + \ldots  + \tilde{a}_{n_{\textrm{cyc}-1}} z + \tilde{a}_{n_{\textrm{cyc}}}
\end{equation}
be the characteristic polynomial of $L_{\textrm{max}}$, considered as an induced subgraph. It is immediate from the definition of $L_{\textrm{max}}$ (and using Sachs' theorem) that $\tilde{a}_i = a_i$ for all $i$. Thus, the additional eigenvalues of $\vec{S}$ are the roots of Eq.~\ref{leval} which are the eigenvalues of the disjoint cycles in $L_{\textrm{max}}$ and the result follows.
\end{proof}

\begin{proposition}
Let $\vec{S}$ be a cycle-disjoint signed digraph with $n$ vertices, adjacency matrix $\mathbf{A}$ and eigenvalues $\lambda_1, \ldots, \lambda_n$. Let $\mathbf{b}=\text{diag}(-\epsilon_1,\ldots,-\epsilon_n)$ be a diagonal matrix with $\epsilon_i \ge 0$ for all $i$ and let $\mu_1,\ldots,\mu_n$ be the eigenvalues of the matrix $\mathbf{B} = \mathbf{A} + \mathbf{b}$. Then
$$
	\max_i \textup{Re}(\mu_i) \le \max_i \textup{Re}(\lambda_i)
$$ 
with equality only when $\epsilon_i=0$ for all $i$.
\end{proposition}
\begin{proof}
The proof consists of three parts: (1) reduction of the problem to that of a cycle; (2) proof for the positive cycle case; and (3) proof for the negative cycle case.
 
\emph{Part 1: reduction of the problem to that of a cycle}. The matrix $\mathbf{B} = \mathbf{A} + \mathbf{b}$ may be thought of as the adjacency matrix of a weighted digraph $\vec{P}$ which has the same vertices, edges and edge-signs as $\vec{S}$ with  an additional self-loop at each vertex $v_i \in \vec{P}$ of weight $-\epsilon_i$. Let $v$ be a vertex not participating in a cycle in $\vec{S}$. Any linear subgraph $L$ of $\vec{P}$ containing $v$ can only do so via the self-loop at $v$ and hence $v$ must be disjoint from all other vertices in $L$. Thus, removing all edges in $\vec{P}$ which do not participate in a cycle in $\vec{S}$, except the weighted self-loops, creates a new graph $\vec{Q}$ with the same linear subgraphs as $\vec{P}$ and thus the same characteristic polynomial as $\vec{P}$ by Sachs' theorem. The graph $\vec{Q}$ consists of the disjoint union of the perturbed cycles of $\vec{S}$ and $(n-n_{\textrm{cyc}})$ isolated vertices each with a self-loop of weight $-\epsilon_i$ for some $i$. Each of the isolated vertices contributes the eigenvalue $-\epsilon_i < 0$ for some $i$ to the spectrum of $\vec{Q}$. The remainder of the spectrum is determined by the disjoint union of perturbed cycles in $\vec{Q}$. Since the spectrum of a disjoint union of graphs is the union of the spectra of each of the components, it is therefore sufficient to prove the proposition for $\vec{S}$ a cycle. In particular, if $\vec{S}$ is acyclic then the spectra of $\vec{P}$ is just the set $\{-\epsilon_i \, \vert \, i = 1 \ldots n \}$ which has maximal real-part $-\textrm{min} \, \epsilon_i < 0$ as stated in the main text.

\emph{Part 2: proof for the positive cycle case}. Let $\vec{C}_n^+$ be a positive cycle with $n$ vertices and adjacency matrix $\mathbf{A}$. The characteristic polynomial of $\vec{C}_n^+$ is $z^n - 1$ and the eigenvalues of $\vec{C}_n^+$ are therefore the $n$th roots of unity, which have maximal real-part $1$ for any $n$. The adjacency matrix of a perturbed positive cycle is $\mathbf{A} + \textup{diag}(-\epsilon_1,\ldots,-\epsilon_n)$, which has characteristic polynomial (cf.~Eq.~\ref{Sach})
$$
	p_n^+(z) = (z+\epsilon_1)\cdot (z + \epsilon_2) \cdot \ldots \cdot(z+\epsilon_n) - 1 = \prod_{i = 1}^{n} (z  + \epsilon_i) - 1.
$$
Thus, we need to prove that every root $\lambda$ of $p_n^+$ satisfies $\textup{Re}(\lambda) < 1$ when at least one $\epsilon_i \in \mathbb{R}$ is nonzero. From now on we shall assume, without loss of generality, that $\epsilon_1 > 0$.

We shall use Rouch\'e's theorem, a well-known theorem in complex analysis for locating the roots of functions. For a proof of Rouch\'e's theorem see \cite{remmert}.

\begin{theorem}[Rouch\'e]
Let $f$ and $g$ be holomorphic functions in a domain $R \subset \mathbb{C}$. Let $D \subset R$ be a bounded subset such that its boundary $\partial D$ is a simple closed curve nullhomologous inside $R$. If
\begin{equation}\label{Rouche1}
	|f(z) - g(z)| < |f(z)| \text{ for all } z \in \partial D
\end{equation}
then $f$ and $g$ have the same number of zeros inside $D$.
\end{theorem}

To apply Rouch\'e's theorem we take
\begin{eqnarray*}
f(z) &=& \prod_{i=1}^n (z + \epsilon_i) \\
g(z) &=& p_n^{+}(z)
\end{eqnarray*}
and we will define $D$ in a moment. It is immediate that $|f(z) - g(z)|=1$ for all $z \in \mathbb{C}$. Let $\epsilon_i$ be the ball of radius 1 centered at $-\epsilon_i$ for each $i$ and $\mathscr{B} = \bigcup_i \epsilon_i$ be the union of the balls. Then
\begin{equation}\label{Ineq1}
	|f(z)| = \prod_{i=1}^n |z + \epsilon_i| > 1 \quad \text{if } z \not \in \mathscr{B}.
\end{equation}
Now $f(t)$ with $t \in [0,1]$ is a strictly increasing real function with $f(1)=\prod_{i=1}^n (1 + \epsilon_i) \ge 1+\epsilon_1 > 1$ so there exists $0 \le t_0 < 1$ with $f(t_0) > 1$. Indeed
\begin{equation}\label{Ineq2}
|f(z)| > 1 \quad \text{for all $z \in \mathbb{C}$ with $\textup{Re}(z) = t_0$.}
\end{equation}
To see this, let $z\in \mathbb{C}$ with $\textup{Re}(z)=t_0$ and $\textup{Im}(z)=y$. Then
$$
	|z + \epsilon_i| = \sqrt{(t_0+\epsilon_i)^2 + y^2} \ge \sqrt{(t_0+\epsilon_i)^2} = |t_0 + \epsilon_i| \quad
$$
for all $i=1,\dots,n$ and thus $|f(z)| \ge |f(t_0)| > 1$.\\
Finally, let $a>1$ and $b < \min_i \{-\epsilon_i - 1\}$ and define 
$$
	D = \{z \in \mathbb{C} \;|\; b \le \textup{Re}(z) \le t_0 \textup{ and} -a \le \textup{Im}(z) \le a\}.
$$ 
Then for all $z \in \partial D$ we have either $\text{Re}(z)=t_0$ or $z \not \in \mathscr{B}$ so $|f(z)| > 1 = |f(z)-g(z)|$ by Eq.~\ref{Ineq1} and Eq.~\ref{Ineq2}. Therefore Rouch\'e's Theorem gives that $f(z)$ and $g(z)$ both have all their roots inside $D$. In particular, any root $\lambda$ of $g(z) = p_n^{+}(z)$ satisfies $\textup{Re}(\lambda) < t_0 < 1$ and this proves the positive case. 

\textbf{Remark 1.} Note that, in addition, any root of $p_n^{+}(z)$ with positive real-part must lie \emph{strictly} inside the unit circle. To see this, observe that if $|z|=1$ then $|f(z)|=\prod_i|z+\epsilon_i| \ge |z+\epsilon_1| > 1$ so $z$ is not a root of $p_n^{+}(z)$ by the same argument. We shall make use of this observation in a moment.

\textbf{Remark 2.} The positive case may also be proven by application of a modification of Ger\v sgorin's disc theorem due to Brualdi (see Theorem 6.4.18 in \cite{horn}).

\emph{Part 3: proof for the negative cycle case}. Let $\vec{C}_n^-$ be a negative cycle with $n$ vertices and adjacency matrix $\mathbf{A}$. The characteristic polynomial of $\vec{C}_n^-$ is $z^n + 1$ and the eigenvalues of $\vec{C}_n^-$ are therefore the $n$th roots of $-1$, which have maximal real-part $\cos(\pi/n)$. The adjacency matrix of a perturbed negative cycle is $\mathbf{A} + \textup{diag}(-\epsilon_1,\ldots,-\epsilon_n)$, which has characteristic polynomial (cf.~Eq.~\ref{Sach})
$$
	p_n^-(z) = (z+\epsilon_1)\cdot (z + \epsilon_2) \cdot \ldots \cdot(z+\epsilon_n) + 1 = \prod_{i = 1}^{n} (z  + \epsilon_i) + 1.
$$
Thus, we need to prove that every root $\lambda$ of $p_n^-$ satisfies $\textup{Re}(\lambda) < \cos(\pi/n)$, when at least one $\epsilon_i \in \mathbb{R}$ is nonzero. Again assume, without loss of generality, that $\epsilon_1 > 0$.

We first note that by exactly the same argument as the positive case, we can prove that all the roots $\lambda$ of $p_n^-(z)$ satisfy $\textrm{Re} (\lambda) < 1$ and (by Remark 1) that any root of $p_n^{-}(z)$ with positive real-part must lie strictly inside the unit circle. However, in the negative case this bound is not sufficiently tight to prove the result since the magnitude of the maximal real-part of the eigenvalues depends upon the length of the cycle. In fact, we require the tightest possible bound and the proof in the negative case is correspondingly more involved than that of the positive case. 

We shall progress as before. However this time we use a strengthened version of Rouch\'e's Theorem \cite[p.~390]{remmert} in which the inequality in Eq.~\ref{Rouche1} is replaced by the inequality 
\begin{equation}\label{Estermann}
	|f(z)-g(z)|<|f(z)|+|g(z)| \text{ for all } z \in \partial D.
\end{equation}
In this case, we use the functions 
\begin{eqnarray*}
	f(z) &=& z^n + 1, \\
	g(z) &=& p_n^-(z).
\end{eqnarray*}
and the region
$$
	D_\theta = \{ z=re^{i\varphi} \in \mathbb{C} \;|\; 0\le r \le 1, -\theta \le \varphi \le \theta \}
$$
where $0< \theta < \pi/n$. In particular, we shall prove that 
$$
	|f(z)-g(z)|<|f(z)|+|g(z)| \ \text{ for all } z \in \partial D_\theta\,
$$
from which it follows that $p_n^-(z)$ has the same number of roots that $f(z)$ in $D_\theta$. Since, by construction, $f(z)$ has no roots in $D_\theta$ this implies that all roots $\lambda$ of $p_n^-(z)$ with positive real-part must lie strictly in the unit circle excluding the region $D_\theta$ for all $0< \theta < \pi/n$ and, in particular, that each root $\lambda$ of $p_n^-(z)$ has real-part less than $\cos(\pi/n)$.

First note that in general
$$\begin{array}{rcl}
	|f(z)| + |g(z)| &=&   |f(z)| + |g(z)-f(z)+f(z)|\\
					&\ge& |f(z)| + |g(z)-f(z)|-|f(z)|\\ 
					&=&   |g(z)-f(z)|
\end{array}$$
so the non-strict inequality holds for every $z \in \mathbb{C}$ (observe that equality holds, for instance, for any root of either $f$ or $g$). We therefore only need to demonstrate that 
\begin{equation} \label{equaleqn}
|f(z)-g(z)| \neq |f(z)|+|g(z)| \ \text{ for all } z \in \partial D_\theta
\end{equation}
and the result is proven. To do so, we make use of the following two lemmas, whose proofs we leave to the end.

\begin{lemma}\label{lemma1}
Let $u, v \in \mathbb{C}$. Then 
\begin{equation}\label{uvequal}
	|u| = |u + v| + |v|
\end{equation}
if and only if either $u = v = 0$ or $v = \alpha \, u$ with $0\le \alpha \le -1$.
\end{lemma}
Consider the (open) upper and lower half-planes
\begin{eqnarray*}
	H^+ &=& \{ z \in \mathbb{C} \;|\; \textup{Im}(z) > 0\},\\
	H^- &=& \{ z \in \mathbb{C} \;|\; \textup{Im}(z) < 0\},
\end{eqnarray*}
and the four (open) quadrants
$$Q_k = \{ z = re^{\varphi} \in \mathbb{C} \;|\; r > 0, \frac{(k-1)\pi}{2} < \varphi < \frac{k\pi}{2}\}$$ 
for $k = 1 \ldots 4$. Additionally, for a nonzero complex number $w$ write $\textup{Arg}(w)$ for the unique $\varphi \in (-\pi, \pi]$ such that $w = |w|e^{i\varphi}$.

\begin{lemma}\label{Lemma2} Let $b \in \mathbb{R}^+$.
\begin{enumerate}
\item If $z \in Q_1$ then $|z+b| > |z| > 0$ and $0 < \textup{Arg}(z+b) < \textup{Arg}(z)$.
\item Suppose that $w_1, w_2 \in H^+$ satisfy $|w_1| > |w_2|$ and $\textup{Arg}(w_1) < \textup{Arg}(w_2)$. Then either $w_1-w_2\in H^+$ or $w_1-w_2 \in Q_4 \cup \mathbb{R}^+$.
\end{enumerate}
\end{lemma}

We now apply Lemma \ref{lemma1} and Lemma \ref{Lemma2} to complete the proof. 

Divide $D_\theta$ into three regions:
\begin{eqnarray*}
	D_\theta^+ = D_\theta \cap H^+,\ D_\theta^- = D_\theta \cap H^- \text{ and} \
	D_\theta \cap \mathbb{R}.
\end{eqnarray*}
Let $z \in D_\theta^+$. To make use of Lemma \ref{lemma1} set $u = s_{n-1} z^{n-1} + \ldots + s_1 z + s_0$ and $v= z^n + 1$. We argue by contradiction. Suppose that $z$ does not satisfy Eq.~\ref{equaleqn}, that is, in terms of $u$ and $v$
\begin{equation}
	|u| = |u + v| + |v|.
\end{equation}
Therefore by Lemma \ref{lemma1} $u$ and $v$ lie on a line through the origin. We shall prove that $v \in Q_1$ and $u \not \in Q_3$ and hence arrive at a contradiction.

We know that for $z \in D_\theta$, $0 < \textup{Arg}(z) \le \theta < \pi/n$ and $|z| \le 1$ hence $0 < \textup{Arg}(z^n) < \pi$ and $|z^n| \le 1$. Consequently $v = z^n+1 \in Q_1$. On the other hand, consider 
$$u = s_{n-1} z^{n-1} + \ldots + s_1 z + s_0 = \prod_{i = 1}^n (z + b_i) - z^n.$$
Write $w_1 = \prod_{i = 1}^n (z + b_i)$ and $w_2=z^n$. By Lemma \ref{Lemma2}~(1), $|w_1| > |w_2|$ and $0 < \textup{Arg}(w_1) < \textup{Arg}(w_2)$ since at least $b_1>0$. By Lemma \ref{Lemma2}~(2), $u=w_1 - w_2$ lies in either $H^+$ or $Q_4\cup\mathbb{R}^+$ and thus $u \not \in Q_3$.

If $z \in D_\theta^-$, we apply exactly the same argument to the complex conjugate $\overline{z} \in D_\theta^+$ to conclude that $\overline{u}$ and $\overline{v}$ do not lie in a line through the origin, therefore neither do $u$ and $v$, again contradicting Lemma \ref{lemma1}.

Finally, if $z \in D_\theta \cap \mathbb{R} = [0,1]$ then $u$ and $v$ are both positive real and hence do not satisfy Lemma \ref{lemma1} and this completes the proof.
\end{proof}

\begin{proof}[Proof of Lemma \ref{lemma1}.]
Write $u = a + bi$, $v=c + di$. Then $|u+v| = |v| - |u|$ means
$$
	\sqrt{(a+c)^2 + (b+d)^2} = \sqrt{c^2 + d^2} - \sqrt{a^2 + b^2}
$$
which implies that (squaring and simplifying)
$$
	ac+bd = \sqrt{(a^2+b^2)(c^2+d^2)}.
$$
Squaring and simplifying again we obtain 
$$
(ad-bc)^2=0 \textup{, that is, } ad = bc\,.
$$
If $u=0$ then Eq.~\ref{uvequal} implies $v=0$ . If $u \neq 0$ then either $\alpha=d/b$ or $\alpha=c/a$ is well-defined and satisfies $v =\alpha \, u$. In addition,
$$
	|u| = |u+v|+|v| = |u+\alpha u| + |\alpha u|  \ \Rightarrow\ 1 = |1+\alpha|+|\alpha|
$$
and a case study shows that $\alpha\le 0$ and $1+\alpha\ge 0$, that is, $-1\le \alpha \le 0$. One finally checks that, for such an $\alpha$, $v = \alpha \, u$ satisfies Eq.~\ref{uvequal}.
\end{proof}

\vspace{-2ex}

\begin{proof}[Proof of Lemma \ref{Lemma2}.]
	(1) Let $z=x+iy$ with $x, y >0$. Then
	$$ |z+b| = \sqrt{(x+b)^2 + y^2} > \sqrt{x^2+y^2} = |z|$$
	since $x+b > x > 0$.
	Recall that $\arctan$ is a strictly increasing function. Thus
	$$ \textup{Arg}(z+b) = \arctan\left(\frac{y}{x+b}\right) < \arctan\left(\frac{y}{x}\right) = \textup{Arg}(z).$$
	(2) Let $w_1 = x_1 + i y_1$ and $w_2 = x_2 + i y_2$. If $y_1 > y_2$ then $\textup{Im}(w_1-w_2) = y_1 - y_2 > 0$ so $w_1-w_2 \in H^+$. Suppose that $y_2 \ge y_1 > 0$ (the reader should draw a picture at this stage to convince themselves). Since $|w_1| > |w_2|$ we have 
	$x_1^2 - x_2^2 > y_2^2 - y_1^2 \ge 0$, that is, $x_1^2 > x_2^2$ or, equivalently, $|x_1| > |x_2|$. Then either (a) $x_1 > |x_2|$ or (b) $x_1 < -|x_2|$. The latter is impossible: if $x_2 \ge 0$ then $0 < \textup{Arg}(w_2) \le \pi/2$ but $x_1 < - x_2 \le 0$ so $\textup{Arg}(w_1) > \pi/2$; if $x_2 < 0$ then $x_1 < x_2 < 0$ and hence 
	$$ \frac{1}{x_2} < \frac{1}{x_1} \Rightarrow \frac{y_2}{x_2} < \frac{y_1}{x_1} \Rightarrow \textup{Arg}(w_2) < \textup{Arg}(w_1).$$
In either case the condition $\textup{Arg}(w_1) < \textup{Arg}(w_2)$ is contradicted. So we must have (a)
$x_1 > |x_2|$, that is, $x_1-x_2>0$ and 
therefore $$w_1-w_2 = (x_1-x_2)+(y_1-y_2)i \in Q_4 \cup \mathbb{R}^+.\qedhere$$
\end{proof}

\section{Dissipation-induced instability}\label{appendixB}
\begin{figure}[t!] 
\begin{center}
\includegraphics[width=0.30\textwidth]{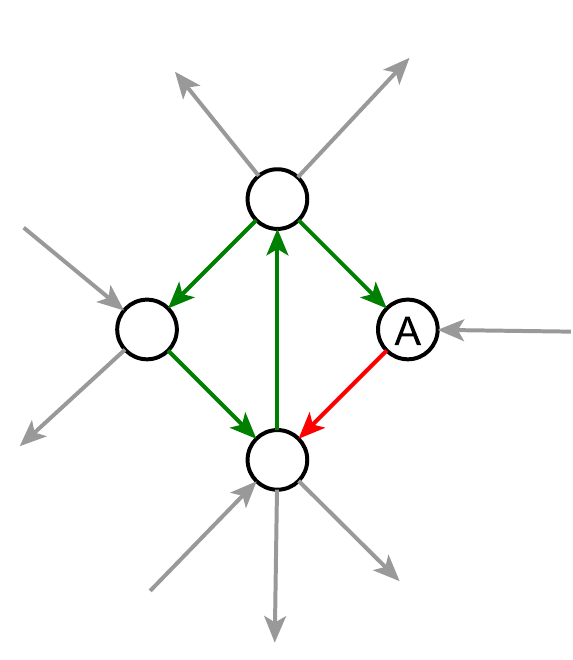}
\caption{\label{examplegraph}\small{A graph which exhibits dissipation-induced instability. Ghost edges show how this graph may be embedded in a larger graph.}}
\end{center}
\end{figure}

Occasionally in our model positive and negative cycles will intersect in a locally symmetric manner such that their contributions to the graph spectrum cancel each other out. In these cases, disparate decay rates may act to break the symmetry, giving rise to `dissipation-induced instabilities' \cite{krechetnikov}. A simple example of when this occurs, consisting of a positive and a negative cycle arranged back-to-back, is given in Fig.~\ref{examplegraph}. In the absence of dissipation the characteristic polynomial of this system is $p_{\mathbf{A}} = \lambda^4$ and the graph has a zero eigenvalue of multiplicity $4$. However, if a small amount of dissipation $\epsilon$ is, for illustrative purposes, present on vertex `A' then the characteristic polynomial becomes $p_{\mathbf{B}} = \mu^4 + \epsilon \mu^3 - \epsilon$ which has roots $\mu_k = \epsilon^{1/4}\,\textrm{e}^{2\pi i k/4} - \epsilon/4 + O(\epsilon ^{7/4})$ for $k = 1 \ldots 4$ as $\epsilon \to 0$ and therefore $\mu_{\textrm{max}} > 0$ (to see this set $\mu' = \epsilon^{-1/4} \, \mu$ to give the rescaled problem $\mu'^4 + \delta \mu'^3 -1 = 0$, where $\delta = \epsilon^{3/4}$, use the ansatz $\mu' = \mu'_0 + \delta \mu'_1 + \ldots$ and solve in the limit $\delta \to 0$). In this case changes in cyclic structure during the evolutionary process may not necessarily trigger concordant changes in stability. For instance removing either of the two edges connected to vertex `A' in Fig.~\ref{examplegraph} decreases $\Phi$ and $\Psi$ yet increases $\mu_{\textrm{max}}$ (see main text for definitions of $\Phi$, $\Psi$ and $\mu_{\textrm{max}}$). In practice, since they rely on rather specific structural configurations, changes in stability which are discordant with changes in cyclic structure are rare in our model. Nevertheless, this observation highlights the general case: stability is determined by the interplay between the cycles of $\vec{S}(t)$, their sign, and the relative rates of decay.

\begin{thebibliography}{10}%
\makeatletter
\providecommand \@ifxundefined [1]{%
 \ifx #1\undefined \expandafter \@firstoftwo
 \else \expandafter \@secondoftwo
\fi
}%
\providecommand \@ifnum [1]{%
 \ifnum #1\expandafter \@firstoftwo
 \else \expandafter \@secondoftwo
\fi
}%
\providecommand \enquote [1]{``#1''}%
\providecommand \bibnamefont  [1]{#1}%
\providecommand \bibfnamefont [1]{#1}%
\providecommand \citenamefont [1]{#1}%
\providecommand\href[0]{\@sanitize\@href}%
\providecommand\@href[1]{\endgroup\@@startlink{#1}\endgroup\@@href}%
\providecommand\@@href[1]{#1\@@endlink}%
\providecommand \@sanitize [0]{\begingroup\catcode`\&12\catcode`\#12\relax}%
\@ifxundefined \pdfoutput {\@firstoftwo}{%
 \@ifnum{\z@=\pdfoutput}{\@firstoftwo}{\@secondoftwo}%
}{%
 \providecommand\@@startlink[1]{\leavevmode}%
 \providecommand\@@endlink[0]{}%
}{%
 \providecommand\@@startlink[1]{%
  \leavevmode
  \pdfstartlink
   attr{/Border[0 0 1 ]/H/I/C[0 1 1]}%
   user{/Subtype/Link/A<</Type/Action/S/URI/URI(#1)>>}%
  \relax
 }%
 \providecommand\@@endlink[0]{\pdfendlink}%
}%
\providecommand \url  [0]{\begingroup\@sanitize \@url }%
\providecommand \@url [1]{\endgroup\@href {#1}{\urlprefix}}%
\providecommand \urlprefix [0]{URL }%
\providecommand \Eprint[0]{\href }%
\@ifxundefined \urlstyle {%
  \providecommand \doi [1]{doi:\discretionary{}{}{}#1}%
}{%
  \providecommand \doi [0]{doi:\discretionary{}{}{}\begingroup
  \urlstyle{rm}\Url }%
}%
\providecommand \doibase [0]{http://dx.doi.org/}%
\providecommand \Doi[1]{\href{\doibase#1}}%
\providecommand \bibAnnote [3]{%
  \BibitemShut{#1}%
  \begin{quotation}\noindent
    \textsc{Key:}\ #2\\\textsc{Annotation:}\ #3%
  \end{quotation}%
}%
\providecommand \bibAnnoteFile [2]{%
  \IfFileExists{#2}{\bibAnnote {#1} {#2} {\input{#2}}}{}%
}%
\providecommand \typeout [0]{\immediate \write \m@ne }%
\providecommand \selectlanguage [0]{\@gobble}%
\providecommand \bibinfo [0]{\@secondoftwo}%
\providecommand \bibfield [0]{\@secondoftwo}%
\providecommand \translation [1]{[#1]}%
\providecommand \BibitemOpen[0]{}%
\providecommand \bibitemStop [0]{}%
\providecommand \bibitemNoStop [0]{.\EOS\space}%
\providecommand \EOS [0]{\spacefactor3000\relax}%
\providecommand \BibitemShut [1]{\csname bibitem#1\endcsname}%
\bibitem{REVTEX41Control}%
  \BibitemOpen
  %
  \bibAnnoteFile{NoStop}{REVTEX41Control}%
\bibitem{apsrev41Control}%
  \BibitemOpen
  \bibfield{author}{%
  \bibinfo {author} {\bibnamefont{08}}}%
   (\bibinfo {year} {1})%
  \bibAnnoteFile{NoStop}{apsrev41Control}%
\bibitem{albert}%
  \BibitemOpen
  \bibfield{author}{%
  \bibinfo {author} {\bibfnamefont{R.}~\bibnamefont{Albert}}\ and\ \bibinfo
  {author} {\bibfnamefont{A.-L.}\ \bibnamefont{Barab\'{a}si}},\ }%
  \bibfield{journal}{%
  \bibinfo {journal} {Rev. Mod. Phys.}\ }%
  \textbf{\bibinfo {volume} {74}},\ \bibinfo {pages} {47} (\bibinfo {year}
  {2002})%
  \bibAnnoteFile{NoStop}{albert}%
\bibitem{newman}%
  \BibitemOpen
  \bibfield{author}{%
  \bibinfo {author} {\bibfnamefont{M.~E.~J.}\ \bibnamefont{Newman}},\ }%
  \bibfield{journal}{%
  \bibinfo {journal} {SIAM Rev.}\ }%
  \textbf{\bibinfo {volume} {45}},\ \bibinfo {pages} {167} (\bibinfo {year}
  {2003})%
  \bibAnnoteFile{NoStop}{newman}%
\bibitem{may}%
  \BibitemOpen
  \bibfield{author}{%
  \bibinfo {author} {\bibfnamefont{R.~M.}\ \bibnamefont{May}},\ }%
  \bibfield{journal}{%
  \bibinfo {journal} {Nature}\ }%
  \textbf{\bibinfo {volume} {238}},\ \bibinfo {pages} {413} (\bibinfo {year}
  {1972})%
  \bibAnnoteFile{NoStop}{may}%
\bibitem{thomas}%
  \BibitemOpen
  \bibfield{author}{%
  \bibinfo {author} {\bibfnamefont{R.}~\bibnamefont{Thomas}}\ and\ \bibinfo
  {author} {\bibfnamefont{R.}~\bibnamefont{D'Ari}},\ }%
  \emph{\bibinfo {title} {Biological feedback}}\ (\bibinfo {publisher} {CRC
  Press},\ \bibinfo {year} {1990})%
  \bibAnnoteFile{NoStop}{thomas}%
\bibitem{batty}%
  \BibitemOpen
  \bibfield{author}{%
  \bibinfo {author} {\bibfnamefont{M.}~\bibnamefont{Batty}},\ }%
  \bibfield{journal}{%
  \bibinfo {journal} {Nature}\ }%
  \textbf{\bibinfo {volume} {444}},\ \bibinfo {pages} {592} (\bibinfo {year}
  {2006})%
  \bibAnnoteFile{NoStop}{batty}%
\bibitem{luscombe}%
  \BibitemOpen
  \bibfield{author}{%
  \bibinfo {author} {\bibfnamefont{N.~M.}\ \bibnamefont{Luscombe}}, \bibinfo
  {author} {\bibfnamefont{M.}~\bibnamefont{Madan~Babu}}, \bibinfo {author}
  {\bibfnamefont{H.}~\bibnamefont{Yu}}, \bibinfo {author}
  {\bibfnamefont{M.}~\bibnamefont{Snyder}}, \bibinfo {author}
  {\bibfnamefont{S.~A.}\ \bibnamefont{Teichmann}},\ and\ \bibinfo {author}
  {\bibfnamefont{M.}~\bibnamefont{Gerstein}},\ }%
  \bibfield{journal}{%
  \bibinfo {journal} {Nature}\ }%
  \textbf{\bibinfo {volume} {431}},\ \bibinfo {pages} {308} (\bibinfo {year}
  {2004})%
  \bibAnnoteFile{NoStop}{luscombe}%
\bibitem{gautreau}%
  \BibitemOpen
  \bibfield{author}{%
  \bibinfo {author} {\bibfnamefont{A.}~\bibnamefont{Gautreau}}, \bibinfo
  {author} {\bibfnamefont{A.}~\bibnamefont{Barrat}},\ and\ \bibinfo {author}
  {\bibfnamefont{M.}~\bibnamefont{Barth{\'e}lemy}},\ }%
  \bibfield{journal}{%
  \bibinfo {journal} {Proc. Nat. Acad. Sci. USA}\ }%
  \textbf{\bibinfo {volume} {106}},\ \bibinfo {pages} {8847} (\bibinfo {year}
  {2009})%
  \bibAnnoteFile{NoStop}{gautreau}%
\bibitem{gross}%
  \BibitemOpen
  \bibfield{author}{%
  \bibinfo {author} {\bibfnamefont{T.}~\bibnamefont{Gross}}\ and\ \bibinfo
  {author} {\bibfnamefont{B.}~\bibnamefont{Blasius}},\ }%
  \bibfield{journal}{%
  \bibinfo {journal} {J. R. Soc. Interface}\ }%
  \textbf{\bibinfo {volume} {5}},\ \bibinfo {pages} {259} (\bibinfo {year}
  {2008})%
  \bibAnnoteFile{NoStop}{gross}%
\bibitem{beggs}%
  \BibitemOpen
  \bibfield{author}{%
  \bibinfo {author} {\bibfnamefont{J.~M.}\ \bibnamefont{Beggs}}\ and\ \bibinfo
  {author} {\bibfnamefont{D.}~\bibnamefont{Plenz}},\ }%
  \bibfield{journal}{%
  \bibinfo {journal} {J. Neurosci.}\ }%
  \textbf{\bibinfo {volume} {23}},\ \bibinfo {pages} {11167} (\bibinfo {year}
  {2003})%
  \bibAnnoteFile{NoStop}{beggs}%
\bibitem{shmulevich}%
  \BibitemOpen
  \bibfield{author}{%
  \bibinfo {author} {\bibfnamefont{I.}~\bibnamefont{Shmulevich}}, \bibinfo
  {author} {\bibfnamefont{S.~A.}\ \bibnamefont{Kauffman}},\ and\ \bibinfo
  {author} {\bibfnamefont{M.}~\bibnamefont{Aldana}},\ }%
  \bibfield{journal}{%
  \bibinfo {journal} {Proc. Natl. Acad. Sci. USA}\ }%
  \textbf{\bibinfo {volume} {102}},\ \bibinfo {pages} {13439} (\bibinfo {year}
  {2005})%
  \bibAnnoteFile{NoStop}{shmulevich}%
\bibitem{nykter}%
  \BibitemOpen
  \bibfield{author}{%
  \bibinfo {author} {\bibfnamefont{M.}~\bibnamefont{Nykter}}, \bibinfo {author}
  {\bibfnamefont{N.~D.}\ \bibnamefont{Price}}, \bibinfo {author}
  {\bibfnamefont{M.}~\bibnamefont{Aldana}}, \bibinfo {author}
  {\bibfnamefont{S.~A.}\ \bibnamefont{Ramsey}}, \bibinfo {author}
  {\bibfnamefont{S.~A.}\ \bibnamefont{Kauffman}}, \bibinfo {author}
  {\bibfnamefont{L.~E.}\ \bibnamefont{Hood}}, \bibinfo {author}
  {\bibfnamefont{O.}~\bibnamefont{Yli-Harja}},\ and\ \bibinfo {author}
  {\bibfnamefont{I.}~\bibnamefont{Shmulevich}},\ }%
  \bibfield{journal}{%
  \bibinfo {journal} {Proc. Natl. Acad. Sci. USA}\ }%
  \textbf{\bibinfo {volume} {105}},\ \bibinfo {pages} {1897} (\bibinfo {year}
  {2008})%
  \bibAnnoteFile{NoStop}{nykter}%
\bibitem{balleza}%
  \BibitemOpen
  \bibfield{author}{%
  \bibinfo {author} {\bibfnamefont{E.}~\bibnamefont{Balleza}}, \bibinfo
  {author} {\bibfnamefont{E.~R.}\ \bibnamefont{Alvarez-Buylla}}, \bibinfo
  {author} {\bibfnamefont{A.}~\bibnamefont{Chaos}}, \bibinfo {author}
  {\bibfnamefont{S.}~\bibnamefont{Kauffman}}, \bibinfo {author}
  {\bibfnamefont{I.}~\bibnamefont{Shmulevich}},\ and\ \bibinfo {author}
  {\bibfnamefont{M.}~\bibnamefont{Aldana}},\ }%
  \bibfield{journal}{%
  \bibinfo {journal} {PLoS One}\ }%
  \textbf{\bibinfo {volume} {3}},\ \bibinfo {pages} {e2456} (\bibinfo {year}
  {2008})%
  \bibAnnoteFile{NoStop}{balleza}%
\bibitem{bornholdta}%
  \BibitemOpen
  \bibfield{author}{%
  \bibinfo {author} {\bibfnamefont{S.}~\bibnamefont{Bornholdt}}\ and\ \bibinfo
  {author} {\bibfnamefont{K.}~\bibnamefont{Sneppen}},\ }%
  \bibfield{journal}{%
  \bibinfo {journal} {Phys. Rev. Lett.}\ }%
  \textbf{\bibinfo {volume} {81}},\ \bibinfo {pages} {236} (\bibinfo {year}
  {1998})%
  \bibAnnoteFile{NoStop}{bornholdta}%
\bibitem{luque}%
  \BibitemOpen
  \bibfield{author}{%
  \bibinfo {author} {\bibfnamefont{B.}~\bibnamefont{Luque}}, \bibinfo {author}
  {\bibfnamefont{F.~J.}\ \bibnamefont{Ballesteros}},\ and\ \bibinfo {author}
  {\bibfnamefont{E.~M.}\ \bibnamefont{Muro}},\ }%
  \bibfield{journal}{%
  \bibinfo {journal} {Phys. Rev. E}\ }%
  \textbf{\bibinfo {volume} {63}},\ \bibinfo {pages} {51913} (\bibinfo {year}
  {2001})%
  \bibAnnoteFile{NoStop}{luque}%
\bibitem{bornholdt2}%
  \BibitemOpen
  \bibfield{author}{%
  \bibinfo {author} {\bibfnamefont{S.}~\bibnamefont{Bornholdt}}\ and\ \bibinfo
  {author} {\bibfnamefont{T.}~\bibnamefont{R\"{o}hl}},\ }%
  \bibfield{journal}{%
  \bibinfo {journal} {Phys. Rev. E}\ }%
  \textbf{\bibinfo {volume} {67}},\ \bibinfo {pages} {066118} (\bibinfo {year}
  {2003})%
  \bibAnnoteFile{NoStop}{bornholdt2}%
\bibitem{liu}%
  \BibitemOpen
  \bibfield{author}{%
  \bibinfo {author} {\bibfnamefont{M.}~\bibnamefont{Liu}}\ and\ \bibinfo
  {author} {\bibfnamefont{K.~E.}\ \bibnamefont{Bassler}},\ }%
  \bibfield{journal}{%
  \bibinfo {journal} {Phys. Rev. E}\ }%
  \textbf{\bibinfo {volume} {74}},\ \bibinfo {pages} {041910} (\bibinfo {year}
  {2006})%
  \bibAnnoteFile{NoStop}{liu}%
\bibitem{garlaschelli}%
  \BibitemOpen
  \bibfield{author}{%
  \bibinfo {author} {\bibfnamefont{D.}~\bibnamefont{Garlaschelli}}, \bibinfo
  {author} {\bibfnamefont{A.}~\bibnamefont{Capocci}},\ and\ \bibinfo {author}
  {\bibfnamefont{G.}~\bibnamefont{Caldarelli}},\ }%
  \bibfield{journal}{%
  \bibinfo {journal} {Nat. Phys.}\ }%
  \textbf{\bibinfo {volume} {3}},\ \bibinfo {pages} {813} (\bibinfo {year}
  {2007})%
  \bibAnnoteFile{NoStop}{garlaschelli}%
\bibitem{rohlf}%
  \BibitemOpen
  \bibfield{author}{%
  \bibinfo {author} {\bibfnamefont{T.}~\bibnamefont{Rohlf}},\ }%
  \bibfield{journal}{%
  \bibinfo {journal} {Europhys. Lett.}\ }%
  \textbf{\bibinfo {volume} {84}},\ \bibinfo {pages} {10004} (\bibinfo {year}
  {2008})%
  \bibAnnoteFile{NoStop}{rohlf}%
\bibitem{magnasco}%
  \BibitemOpen
  \bibfield{author}{%
  \bibinfo {author} {\bibfnamefont{M.~O.}\ \bibnamefont{Magnasco}}, \bibinfo
  {author} {\bibfnamefont{O.}~\bibnamefont{Piro}},\ and\ \bibinfo {author}
  {\bibfnamefont{G.~A.}\ \bibnamefont{Cecchi}},\ }%
  \bibfield{journal}{%
  \bibinfo {journal} {Phys. Rev. Lett.}\ }%
  \textbf{\bibinfo {volume} {102}},\ \bibinfo {pages} {258102} (\bibinfo {year}
  {2009})%
  \bibAnnoteFile{NoStop}{magnasco}%
\bibitem{meisel}%
  \BibitemOpen
  \bibfield{author}{%
  \bibinfo {author} {\bibfnamefont{C.}~\bibnamefont{Meisel}}\ and\ \bibinfo
  {author} {\bibfnamefont{T.}~\bibnamefont{Gross}},\ }%
  \bibfield{journal}{%
  \bibinfo {journal} {Phys. Rev. E}\ }%
  \textbf{\bibinfo {volume} {80}},\ \bibinfo {pages} {061917} (\bibinfo {year}
  {2009})%
  \bibAnnoteFile{NoStop}{meisel}%
\bibitem{christensen}%
  \BibitemOpen
  \bibfield{author}{%
  \bibinfo {author} {\bibfnamefont{K.}~\bibnamefont{Christensen}}, \bibinfo
  {author} {\bibfnamefont{R.}~\bibnamefont{Donangelo}}, \bibinfo {author}
  {\bibfnamefont{B.}~\bibnamefont{Koiller}},\ and\ \bibinfo {author}
  {\bibfnamefont{K.}~\bibnamefont{Sneppen}},\ }%
  \bibfield{journal}{%
  \bibinfo {journal} {Phys. Rev. Lett.}\ }%
  \textbf{\bibinfo {volume} {81}},\ \bibinfo {pages} {2380} (\bibinfo {year}
  {1998})%
  \bibAnnoteFile{NoStop}{christensen}%
\bibitem{bornholdt}%
  \BibitemOpen
  \bibfield{author}{%
  \bibinfo {author} {\bibfnamefont{S.}~\bibnamefont{Bornholdt}}\ and\ \bibinfo
  {author} {\bibfnamefont{T.}~\bibnamefont{Rohlf}},\ }%
  \bibfield{journal}{%
  \bibinfo {journal} {Phys. Rev. Lett.}\ }%
  \textbf{\bibinfo {volume} {84}},\ \bibinfo {pages} {6114} (\bibinfo {year}
  {2000})%
  \bibAnnoteFile{NoStop}{bornholdt}%
\bibitem{maayan}%
  \BibitemOpen
  \bibfield{author}{%
  \bibinfo {author} {\bibfnamefont{A.}~\bibnamefont{Ma'ayan}}, \bibinfo
  {author} {\bibfnamefont{G.~A.}\ \bibnamefont{Cecchi}}, \bibinfo {author}
  {\bibfnamefont{J.}~\bibnamefont{Wagner}}, \bibinfo {author}
  {\bibfnamefont{A.~R.}\ \bibnamefont{Rao}}, \bibinfo {author}
  {\bibfnamefont{R.}~\bibnamefont{Iyengar}},\ and\ \bibinfo {author}
  {\bibfnamefont{G.}~\bibnamefont{Stolovitzky}},\ }%
  \bibfield{journal}{%
  \bibinfo {journal} {Proc. Natl. Acad. Sci. USA}\ }%
  \textbf{\bibinfo {volume} {105}},\ \bibinfo {pages} {19235} (\bibinfo {year}
  {2008})%
  \bibAnnoteFile{NoStop}{maayan}%
\bibitem{Note1}%
  \BibitemOpen
  \bibinfo {note} {For efficiency in the simulations shown we set the
  edge-inclusion probability $\approx \protect \qopname \relax o{ln}(n) /n$ and
  consider a maximally sparse connected random graph. Qualitatively the same
  results may be achieved for more dense graphs.}%
  \bibAnnoteFile{Stop}{Note1}%
\bibitem{Note2}%
  \BibitemOpen
  \bibinfo {note} {This stability criterion assumes that $\protect \mathbf
  {B}(t)$ is the Jacobian matrix of a dynamical system evaluated at a
  fixed-point}%
  \bibAnnoteFile{NoStop}{Note2}%
\bibitem{barabasi}%
  \BibitemOpen
  \bibfield{author}{%
  \bibinfo {author} {\bibfnamefont{A.-L.}\ \bibnamefont{Barab\'{a}si}}\ and\
  \bibinfo {author} {\bibfnamefont{R.}~\bibnamefont{Albert}},\ }%
  \bibfield{journal}{%
  \bibinfo {journal} {Science}\ }%
  \textbf{\bibinfo {volume} {286}},\ \bibinfo {pages} {509} (\bibinfo {year}
  {1999})%
  \bibAnnoteFile{NoStop}{barabasi}%
\bibitem{Note3}%
  \BibitemOpen
  \bibinfo {note} {That is, after an initial transient `settling-down' period
  ($4 \times 10^6$ time-steps prior to data shown).}%
  \bibAnnoteFile{Stop}{Note3}%
\bibitem{Note4}%
  \BibitemOpen
  \bibinfo {note} {A cycle of length $k$ is a non-intersecting path of length
  $k$ from a vertex back to itself respecting edge directions.}%
  \bibAnnoteFile{Stop}{Note4}%
\bibitem{estrada3}%
  \BibitemOpen
  \bibfield{author}{%
  \bibinfo {author} {\bibfnamefont{E.}~\bibnamefont{Estrada}}\ and\ \bibinfo
  {author} {\bibfnamefont{N.}~\bibnamefont{Hatano}},\ }%
  \bibfield{journal}{%
  \bibinfo {journal} {Chem. Phys. Lett.}\ }%
  \textbf{\bibinfo {volume} {439}},\ \bibinfo {pages} {247} (\bibinfo {year}
  {2007})%
  \bibAnnoteFile{NoStop}{estrada3}%
\bibitem{estrada2}%
  \BibitemOpen
  \bibfield{author}{%
  \bibinfo {author} {\bibfnamefont{E.}~\bibnamefont{Estrada}}\ and\ \bibinfo
  {author} {\bibfnamefont{N.}~\bibnamefont{Hatano}},\ }%
  \bibfield{journal}{%
  \bibinfo {journal} {Lin. Alg. Appl.}\ }%
  \textbf{\bibinfo {volume} {430}},\ \bibinfo {pages} {1886} (\bibinfo {year}
  {2009})%
  \bibAnnoteFile{NoStop}{estrada2}%
\bibitem{Note5}%
  \BibitemOpen
  \bibinfo {note} {By discretizing the data we are asking: how much does
  knowing whether the network is cyclic or not tell us about whether the system
  is stable or not?}%
  \bibAnnoteFile{Stop}{Note5}%
\bibitem{supmaterials}%
  \BibitemOpen
  \enquote{\bibinfo {title} {{EPAPS} {D}ocument {N}o. {XXX}},}\ %
  \bibAnnoteFile{NoStop}{supmaterials}%
\bibitem{Note6}%
  \BibitemOpen
  \bibinfo {note} {A cycle $c$ is positive (negative) if the product of the
  edge signs in $c$ equals $+1$ ($-1$).}%
  \bibAnnoteFile{Stop}{Note6}%
\bibitem{ungar}%
  \BibitemOpen
  \bibfield{author}{%
  \bibinfo {author} {\bibfnamefont{A.}~\bibnamefont{Ungar}},\ }%
  \bibfield{journal}{%
  \bibinfo {journal} {Amer. Math. Month.}\ }%
  \textbf{\bibinfo {volume} {89}},\ \bibinfo {pages} {688} (\bibinfo {year}
  {1982})%
  \bibAnnoteFile{NoStop}{ungar}%

\bibitem{cvetkovic}
D.M. Cvetkovi\'{c}, M.~Doob, and H.~Sachs.
\newblock {\em Spectra of graphs: theory and applications}.
\newblock Academic Press, 1980.

\bibitem{horn}
R.~A. Horn and C.~R. Johnson.
\newblock {\em Matrix analysis}.
\newblock Cambridge University Press, 1990.

\bibitem{remmert}
R.~Remmert.
\newblock {\em Theory of complex functions}.
\newblock Springer-verlag, 1998.

\bibitem{krechetnikov}
R. Krechetnikov and J.~E. Marsden
  \newblock{Rev. Mod. Phys.}
  \textbf{\newblock{79}}(\newblock{2}),\ \newblock{519-553},\ \newblock{2007}.%

\end{thebibliography}
\end{document}